\documentclass[structabstract]{aa}  
\usepackage{graphicx}
\usepackage{natbib}
\usepackage{threeparttable}
\usepackage{txfonts}
\def \msun{\ifmmode{{\rm\ M}_\odot}\else{${\rm\ M}_\odot$}\fi}

\newcommand{\kms}{kms$^{-1}$}                         % Kms-1

\begin{document}
   \title{On the multiplicity of supernovae within host galaxies}

%   \subtitle{Evidence for wide-spread episodic, bursty star formation}

   \author{J. P. Anderson
          \and
          M. Soto
          }

   \institute{Departamento de Astronom\'ia, Universidad de Chile, Casilla 36-D
              \email{anderson@das.uchile.cl}}

   \date{}

% \abstract{}{}{}{}{} 
% 5 {} token are mandatory
 
  \abstract
   {}
   {We investigate the nature of multiple supernova hosting galaxies, and
     the types of events which they produce.}
   {Using all known historical supernovae, we split host galaxies 
  into samples containing single or
     multiple events. These samples are then characterised in
     terms of their relative supernova fractions, and host properties.}
   {In multiple supernova hosts the ratio of type Ia to
     core-collapse events is lower than in single supernova hosts. For
     core-collapse events there is a suggestion that the ratio of types Ibc to type
   II events is higher in multiples than within single supernova hosts. This second
   increase is dominated by an increase in the number of SNIb. Within multiple
   supernova hosts, supernovae of any given type appear to `prefer' to
   explode in galaxies that are host to the same type of SN. We also find
   that multiple SN hosts have higher T-type morphologies.}
   {While our results suffer from low number statistics, we speculate that
their simplest interpretation is that star formation within
     galaxies is generally of an episodic and bursty nature. This leads to the
     supernovae detected within any particular galaxy to be dominated by those with
     progenitors of a specific age, rather than a random selection from
     standard relative supernova rates, as the latter would be expected if
     star formation was of a long-term continuous nature.\\
We further discuss the supernova progenitor and star formation properties
   that may be important for understanding these trends, and also comment on 
   a range of important selection effects within our sample.}


   \maketitle
%
%________________________________________________________________

\section{Introduction}
The extreme brightness of supernovae (SNe) enables their discovery in stellar
populations outside of our own galaxy. Within the last 127 years there have
been almost 6000 SN discoveries within external galaxies (see
e.g. the IAU\footnote{http://www.cbat.eps.harvard.edu/lists/Supernovae.html} 
and Asiago\footnote{http://graspa.oapd.inaf.it/}; \citealt{bar99}, catalogues). The majority of
galaxies within these catalogues have been host to 1 detected SN. However,
within a significant fraction multiple SNe have been
discovered. Here we investigate the nature of single and multiple SN hosting
galaxies and the types of SNe which they produce.\\
\indent The general consensus is of two distinct explosion scenarios which
produce SNe. Type Ia SNe (SNIa henceforth) are believed to be
thermonuclear explosions of accreting white dwarfs, while all other
types are thought to arise from the core-collapse of massive stars (CC SNe). 
CC SNe are further divided into 3 main observational
classes: SNII which show hydrogen in their spectra; SNIb which lack hydrogen
but show helium; and SNIc which lack both hydrogen and helium in their spectra
(see \citealt{fil97} for a review of SN classifications). SNII can be further
sub-divided into IIP, IIL, IIb and IIn dependent on the nature of their
light-curve and spectral features. However, for the current
analysis we do not use these further sub-type classifications.\\
\indent Several hundred galaxies have now been host to multiple observed SNe. Given the
general consensus that a typical star-forming galaxy will produce a SN at a
rate of around 1 event per 100 years, these multiple events are expected
through simple statistical fluctuations. We may also expect certain galaxies
to produce SNe at a higher rate if they are simply producing stars at a higher
rate. The most pertinent
questions would then appear to be: are there more galaxies producing
multiple events than would be expected if drawing randomly from a standard SN
rate? 
What are the characteristics of these multiple SN
producers? Do multiple SN hosts produce the same types of events as
their single SN counterparts?\\
\indent The fact that some galaxies appear particularly efficient at producing SNe
was noted in the early era of SNe study by \cite{zwi38}, and some early
considerations of their hosts were outlined in \cite{kuk65}. \cite{ric88} 
using a sample of 627
single- and 88 multiple-SN hosting galaxies, claimed that some
galaxies go through an epoch of star formation (SF) at a
rate 70 times higher than normal `inert' galaxies. However, this value was
later revised downwards by both \cite{gut90} and \cite{li95} to levels consistent with
statistical fluctuations, once various selections effects were
considered. One particular galaxy: NGC 2770 which
had been host to 3 SNIb was analysed and discussed in detail by
\cite{tho09}. While these authors estimated that there was only a
1.5\%\ chance probability of observing three SNIb in 10 years, 
they assigned this to coincidence rather than some defining
galaxy characteristic. This
second aspect is that which we choose to concentrate on in the current paper:
analysing the relative fractions of SNe as a function of host galaxy SN
multiplicity.\\ 
\indent The paper is organised as follows. First we define the galaxy and SN samples,
then outline how we further divide these for analysis. Our results are
presented in section 3, followed by an analysis and discussion of possible
selection effects which need to be considered in section 4. We present a discussion 
of the implications of these results in
section 5, together with some remarks on those galaxies which have
been host to the highest number of SNe thus far. We conclude in section 6.

\section{The supernova and galaxy sample}
Our initial SN sample is that downloaded from the Asiago catalogue,
with the most recent SN being SN2012cp discovered on the 23rd May 2012
\citep{cox12}. This catalogue contains 5928 SN discoveries. 
We remove all SNe without definitive
classifications as Ia, Ib, Ic or II. We
also remove all SNe without identified hosts. 
After these culls we are left with a sample of 2384 SNe within 2117 host galaxies.
We then search for repeated entries within the host galaxy
listings, and thus separate galaxies into those hosting 1 or multiple events. 
This separation leads to samples of: 1898 SNe within single SN hosts; 
187 galaxies each hosting 2 SNe; 22 galaxies each hosting 3;
7 galaxies each hosting 4; 1 galaxy hosting 5, 1 hosting 6, and 1 galaxy
host to 7 SNe. The combined
multiple sample (all SNe within galaxies hosting 2 or more SNe) has 486 SNe
within 219 galaxies. Once these
samples are formed we then investigate SN ratios and host galaxy
properties within the different samples.

\section{Results}
\subsection{SN ratios}
In table 1 we list the numbers of SNe of different types within 
galaxies host to different numbers of events; for both the full
sample, plus a sub-sample of SNe occurring in SF galaxies. In table 2 we
list the numbers and respective percentages for only the CC SNe, first
listing SNe within galaxies list to different overall numbers of events, then
listing SNe within galaxies host to different numbers of CC events.
\begin{table*}
\centering
\caption{Number of SNe in galaxies which have been host to 1, $\ge$2, then, 2, 3,
and $\ge$4 events. Column 1 gives the number of SNe per host
galaxy. Columns 2, 3, 4, 5, 6 and 7 give the number of SNIa, SNII, SNIb, SNIc,
the total SNIbc (in some cases higher than the sum of SNIb
and SNIc due to some events being classified as `SNIb/c' within
the literature) and the total number of CC SNe respectively, with the percentages of each SN type of the
whole sample, for each particular multiplicity
shown in brackets. First the numbers and percentages are given for the full
sample. Then below the values are listed with all SNe which occurred within
negative T-type host galaxies removed (i.e. SNe in star-forming galaxies).}
\begin{tabular}[t]{ccccccc}
\hline
N SNe per galaxy & NIa(\%) & NII(\%) & NIb(\%) & NIc(\%) & Total NIbc(\%)&Total NCC(\%)  \\
\hline	
\hline
1       & 883(47) & 797(42) &  57(3) & 128(7) & 218(11)& 1015(53) \\
$\geq$2 & 165(34) & 240(49) &  34(7) & 39(8)  & 81(17) & 321(66)  \\      
\hline
2       & 140(37) & 171(46) &  25(7) & 32(9)  & 63(17) & 234(63) \\
3       & 14(21)  & 36(55)  &  7(11) & 7(11)  & 16(24) & 52(79) \\
$\geq$4 & 11(24)  & 33(72)  &  2(4)  & 0(0)   & 2(4)   & 35(76)  \\
\hline
\hline
\textbf{SNe in SF galaxies}&&&&&&\\
1       & 513(37) & 692(49) &  50(4) & 117(8) & 196(14)& 888(63) \\
$\geq$2 & 127(29) & 231(53) &  33(8) & 39(9)  & 78(18) & 309(71)  \\      
\hline
2       & 111(33) & 162(49) &  24(7) & 32(10) & 61(18) & 223(67) \\
3       & 9(15)   & 36(60)  &  7(12) & 7(12)  & 15(25) & 51(85)\\
$\geq$4 & 7(17)   & 33(79)  &  2(5)  & 0(0)   & 2(5)   & 35(83)\\ 
\hline
\hline
\end{tabular}
\end{table*}

\begin{table*}
\centering
\caption{Here we first show the number of CC SNe in galaxies which have been host to 1, $\ge$2, then, 2, 3,
and $\ge$4 SNe. Column 1 gives the number of SNe per host
galaxy. Columns 2, 3, 4, and 5 give the number of SNII, SNIb, SNIc,
and the total SNIbc respectively, with the percentages of each SN type of the
whole CC sample, for each particular multiplicity
shown in brackets. Then below we show the number of CC SNe in galaxies which have been host to 1, $\ge$2, then, 2, 3,
and $\ge$4 CC SNe.}
\begin{tabular}[t]{ccccc}
\hline
N SNe per galaxy & NII(\%) & NIb(\%) & NIc(\%) & Total NIbc(\%)  \\
\hline	
\hline
1       & 797(79) &  57(5) & 128(13) & 218(21)\\
$\geq$2 & 240(75) &  34(11) & 39(12)  & 81(25) \\      
\hline
2       & 171(73) &  25(11) & 32(14)  & 63(27) \\
3       & 36(69)  &  7(13) & 7(13)  & 16(31) \\
$\geq$4 & 33(94)  &  2(6)  & 0(0)   & 2(6)   \\
\hline
\hline
N CC SNe per galaxy & & & &  \\
1       & 846(78) &  65(6) & 143(13) & 243(22)\\
$\geq$2 & 191(77) &  25(10) & 39(10)  & 56(23) \\      
\hline
2       & 135(76) &  19(11) & 18(10)  & 43(24) \\
3       & 37(77)  &  4(8) & 6(13)  & 16(23) \\
$\geq$4 & 19(90)  &  2(10)  & 0(0)   & 2(10)   \\
\hline
\hline
\end{tabular}
\end{table*}
In figure 1 we show the distribution of SN types in single and multiple
SN hosts. We observe that the ratio of SNIa to CC events decreases
significantly in multiple SN hosts. In single SN hosts 
SNIa contribute 47\%\ of the sample, while in multiple SN hosts this
falls to 34\%. The ratio of SNIa to CC events is 0.870$\pm$0.040 in
single SN hosts, and 0.514$\pm$0.049 in multiple SN hosts (errors are
estimated assuming Poisson errors: these are the Poisson statistics 
on the number of SNe within each
sample, with this error then propagated to give a ratio error; all errors
presented hereafter are estimated in the same way). We
calculate that this difference is significant at the 5.7 sigma level.\\ 
\indent SNIa are found in non star-forming elliptical galaxies, while CC events are
usually not (see e.g. \citealt{van05}). Therefore we also investigate how the SNIa to CC
ratio changes when we remove all non star-forming host galaxies. We do this by
removing all SNe which have occurred in galaxies with negative T-type
classifications \citep{dev59}. The resulting distributions of events
are shown in fig. 2. 
The ratio of SNIa to CC events in this sub-sample is 0.578$\pm$0.032 in single
SN hosts, and 0.411$\pm$0.043 in multiple SN hosts. The ratio is therefore
still significantly smaller in multiple SN hosts, now at the at the 3.1 sigma level, assuming Poisson
statistics. In fig. 3 we show how the SNIa to CC ratio changes with SN
multiplicity, with SNe within negative T-type galaxies removed.\\ 
\indent Given that SNIa and CC SNe arise from different
progenitor systems, when discussing changes within the CC SN distribution
we remove the SNIa.
\begin{figure}
\includegraphics[width=9cm]{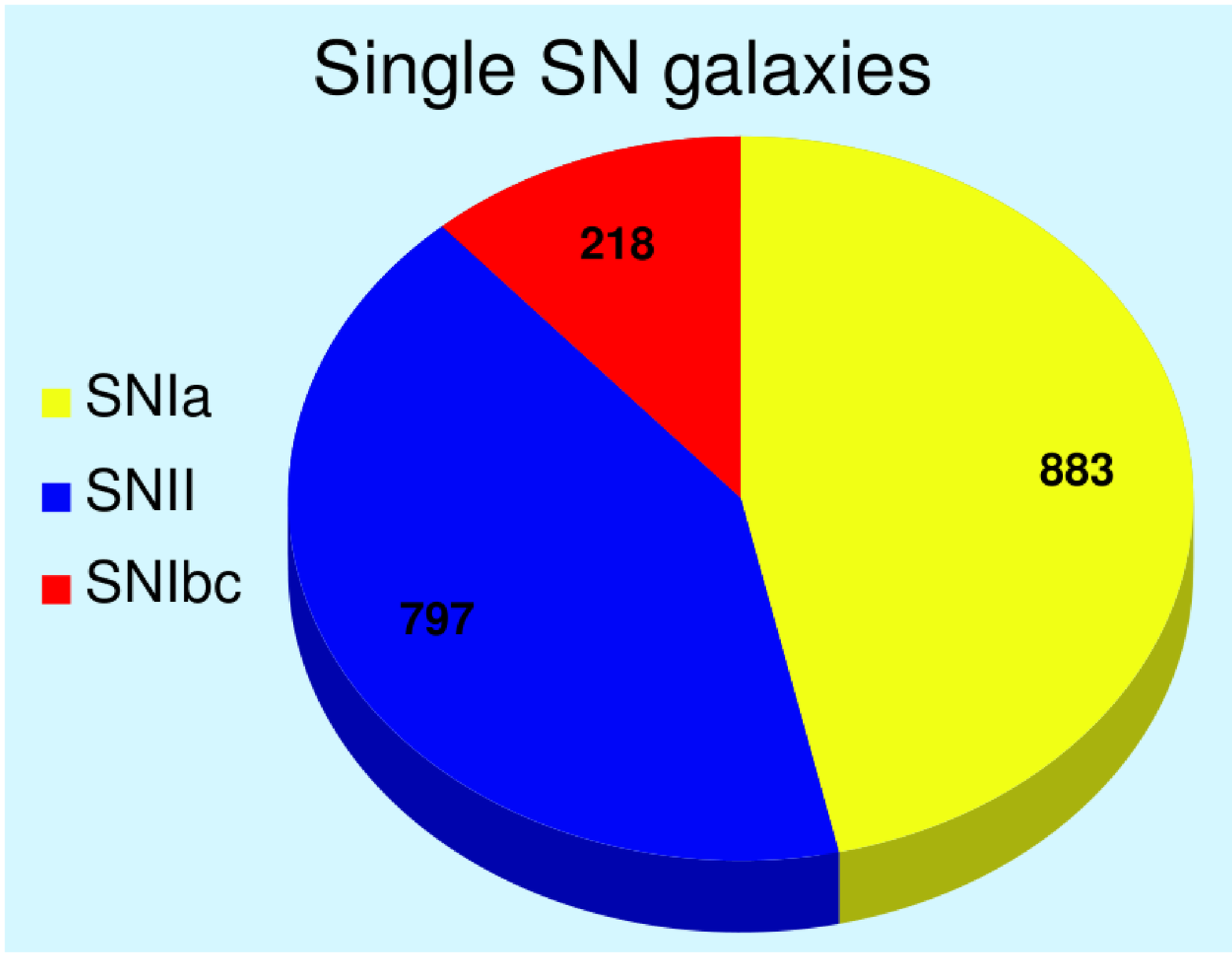}
\includegraphics[width=9cm]{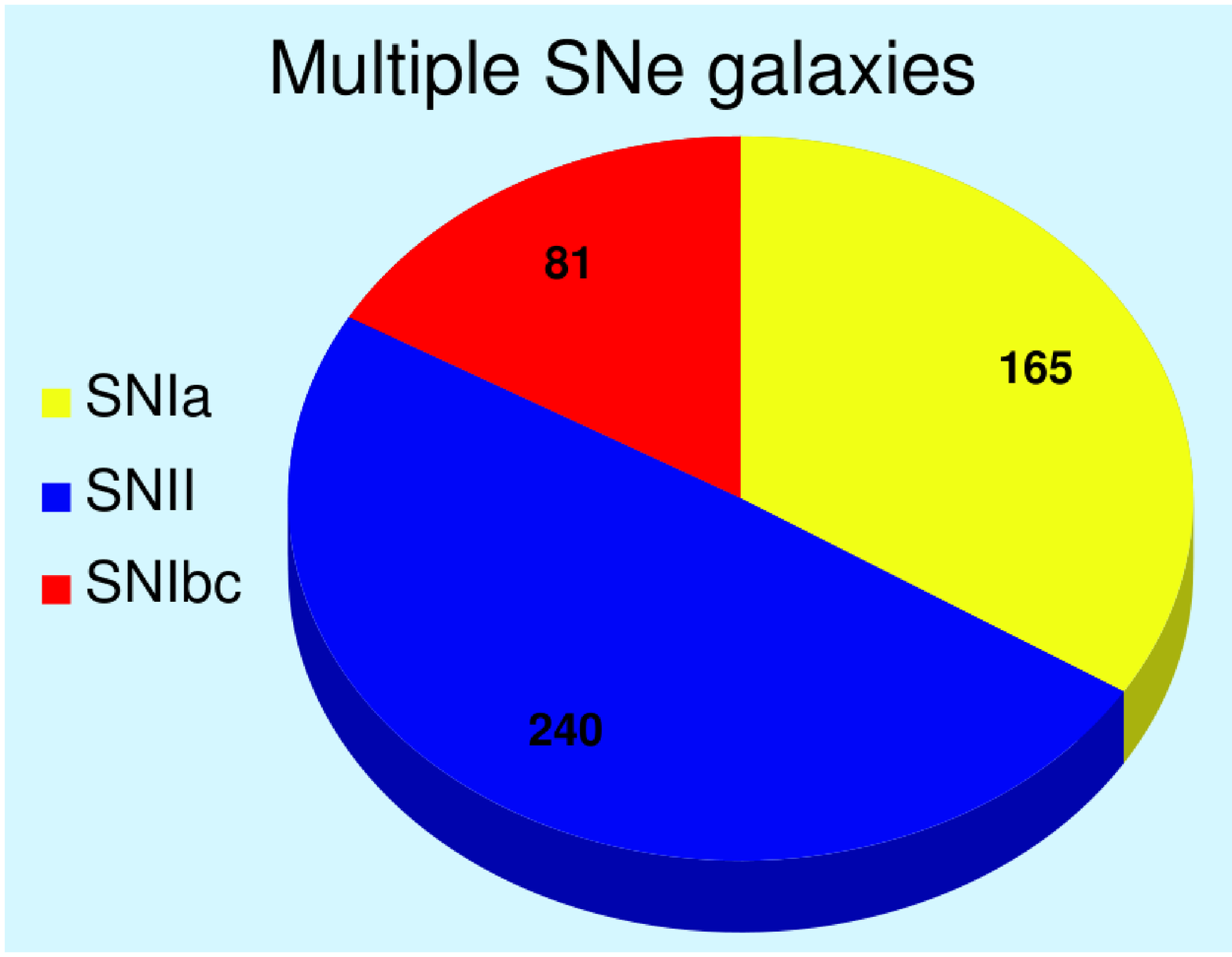}
\caption{Pie charts showing the different SN fractions of events within
single (top) and multiple (bottom) SN hosts.}
\end{figure}
\begin{figure}
\includegraphics[width=9cm]{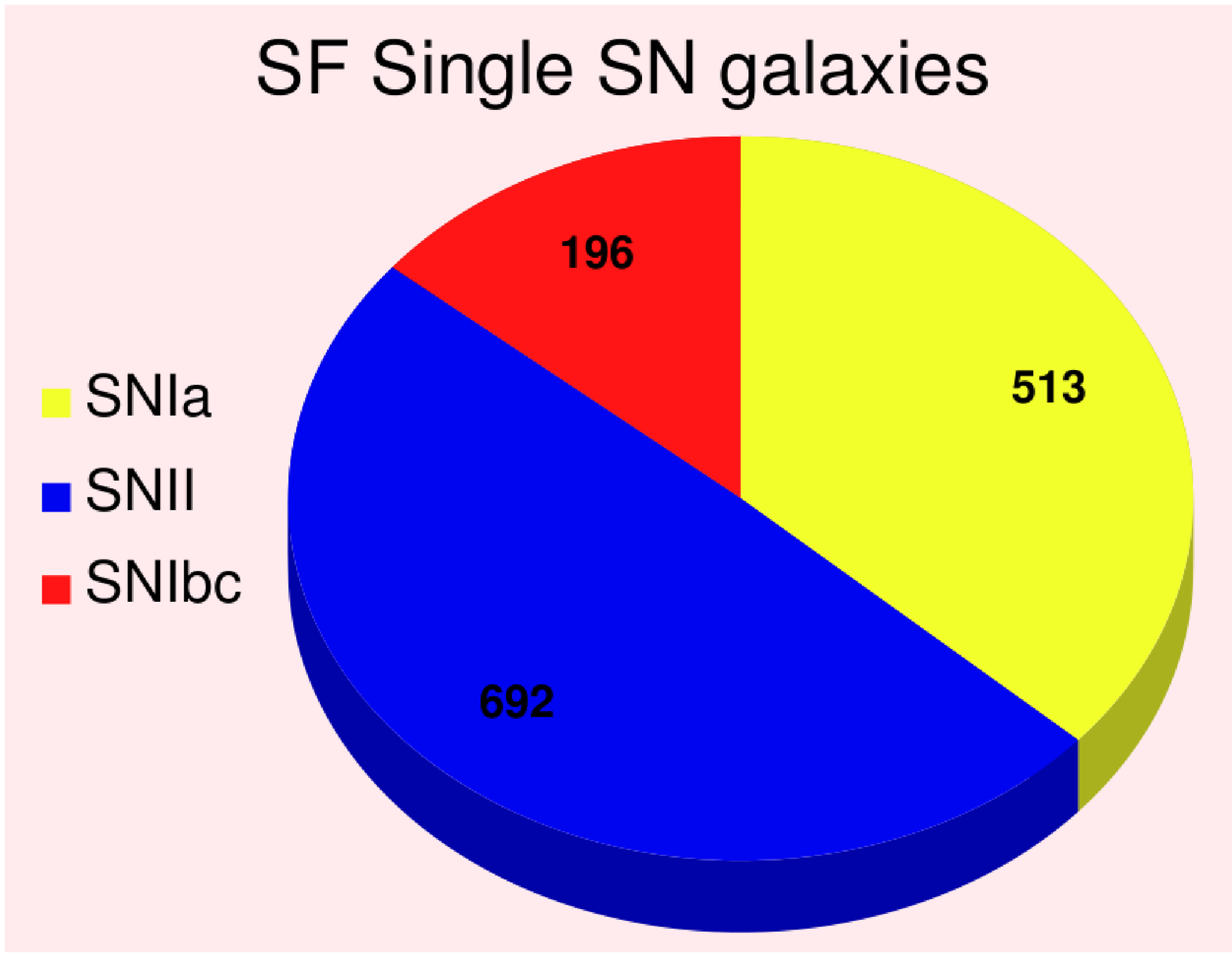}
\includegraphics[width=9cm]{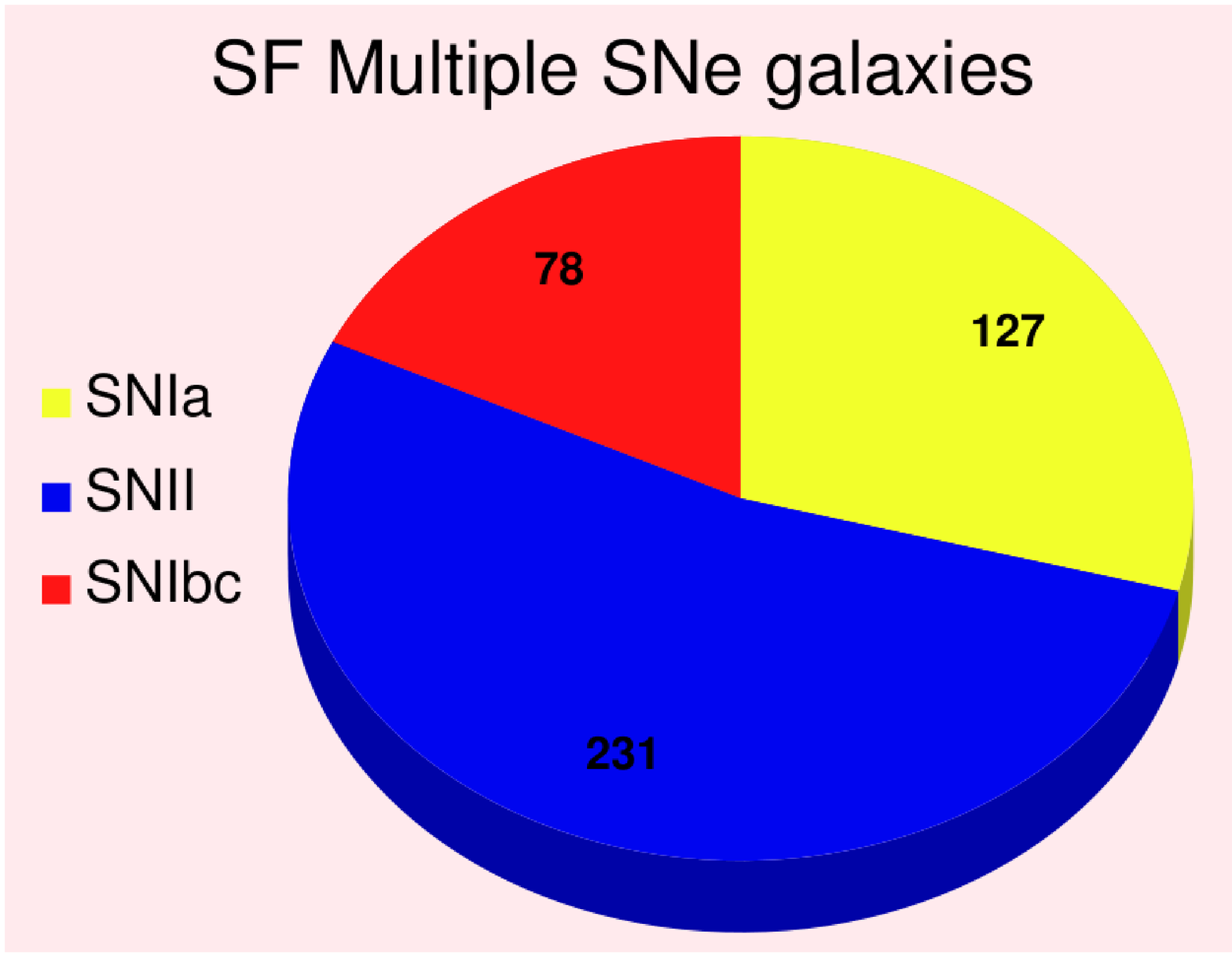}
\caption{Pie charts showing the different SN fractions of events
    within-star forming (i.e. excluding negative T-type host galaxies)
single (top) and multiple (bottom) SN hosts.}
\end{figure}
The relative fractions of CC SNe within single and multiple SN hosts
are shown in fig. 4. We see that the fraction of SNIbc of the total CC sample
increases from 21\%\ in single to 25\%\ in multiple SN hosts. 
The ratio of SNIbc to SNII is 0.274$\pm$0.021 in single SN hosts
and 0.338$\pm$0.047 in multiple SN hosts. Assuming Poisson errors (if the number of SNe within a
distribution is less than 100, then we use \citealt{geh86} to estimate
errors) this difference is
significant at the 1.2 sigma level. We note that when one separates
CC SNe into galaxies that have been host to 1, 2, 3 and $\ge$4 \textit{CC}
events as shown in table 2, this significance is even lower.\\ 
\begin{figure}
\includegraphics[width=9cm]{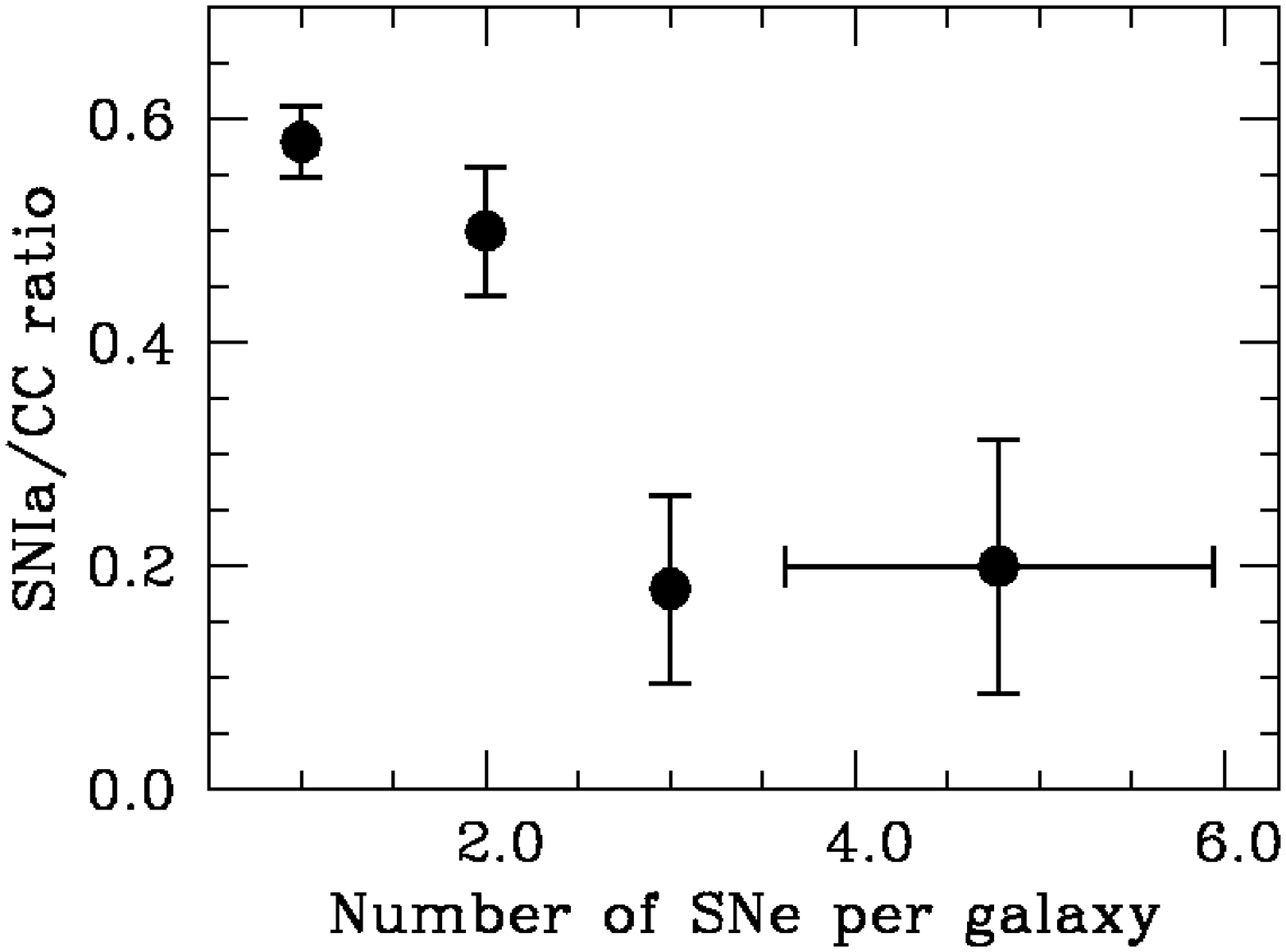}
\caption{Ratio of SNIa to CC events in positive T-type host galaxies as a
function of SN multiplicity. The last point contains all SNe in galaxies
which have been host to $\ge$4 events, with a weighted mean of 4.7
SNe. Error bars on the ratios are
calculated using Poisson statistics (if the number of SNe within a
distribution is less than 100, then we use \citealt{geh86} to estimate
errors). 
The x-axis errors bars on the last point for
$\ge$4
SNe show the standard deviation on the mean number of SNe within each
galaxy within the bin.}
\end{figure}
\begin{figure}
\includegraphics[width=9cm]{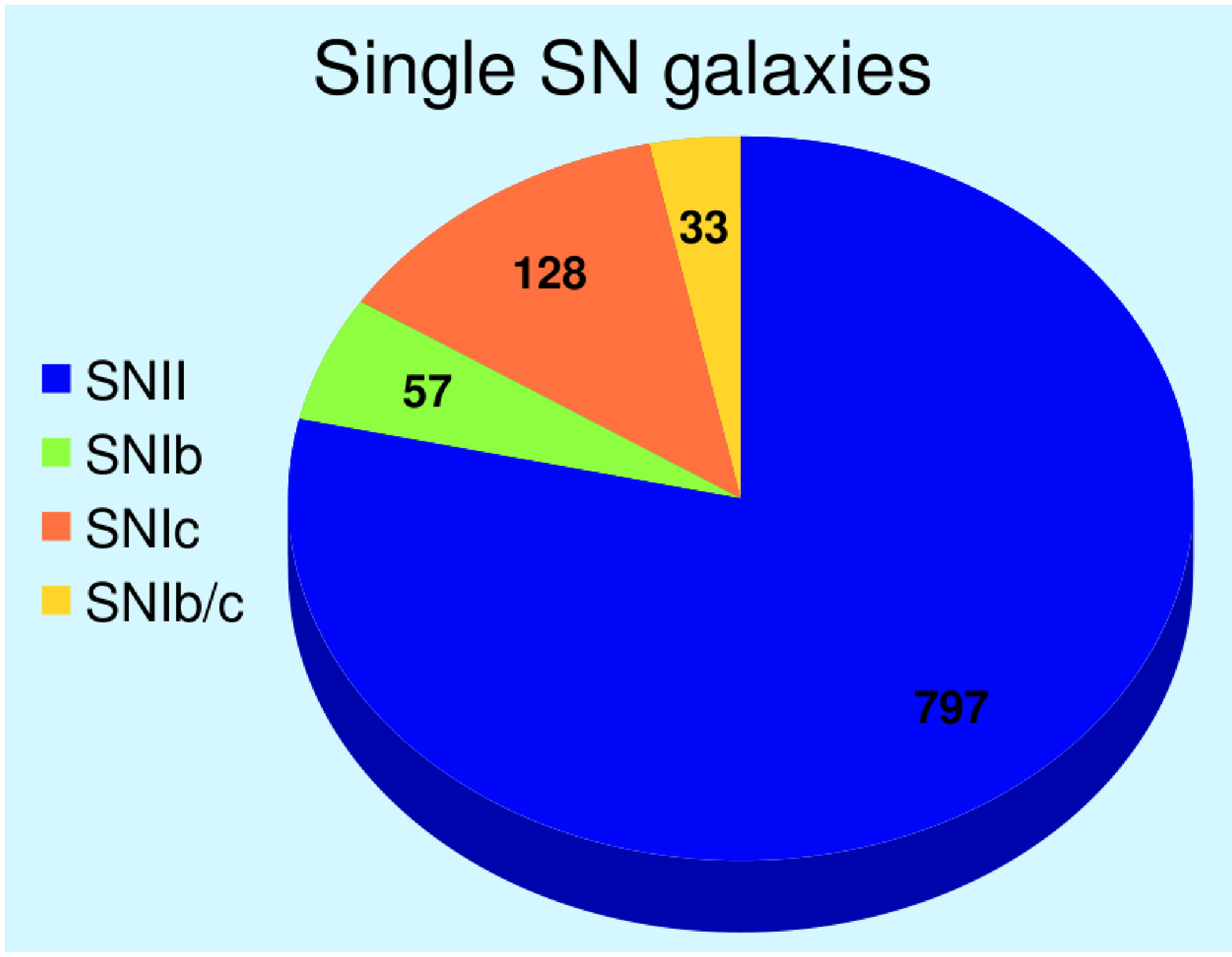}
\includegraphics[width=9cm]{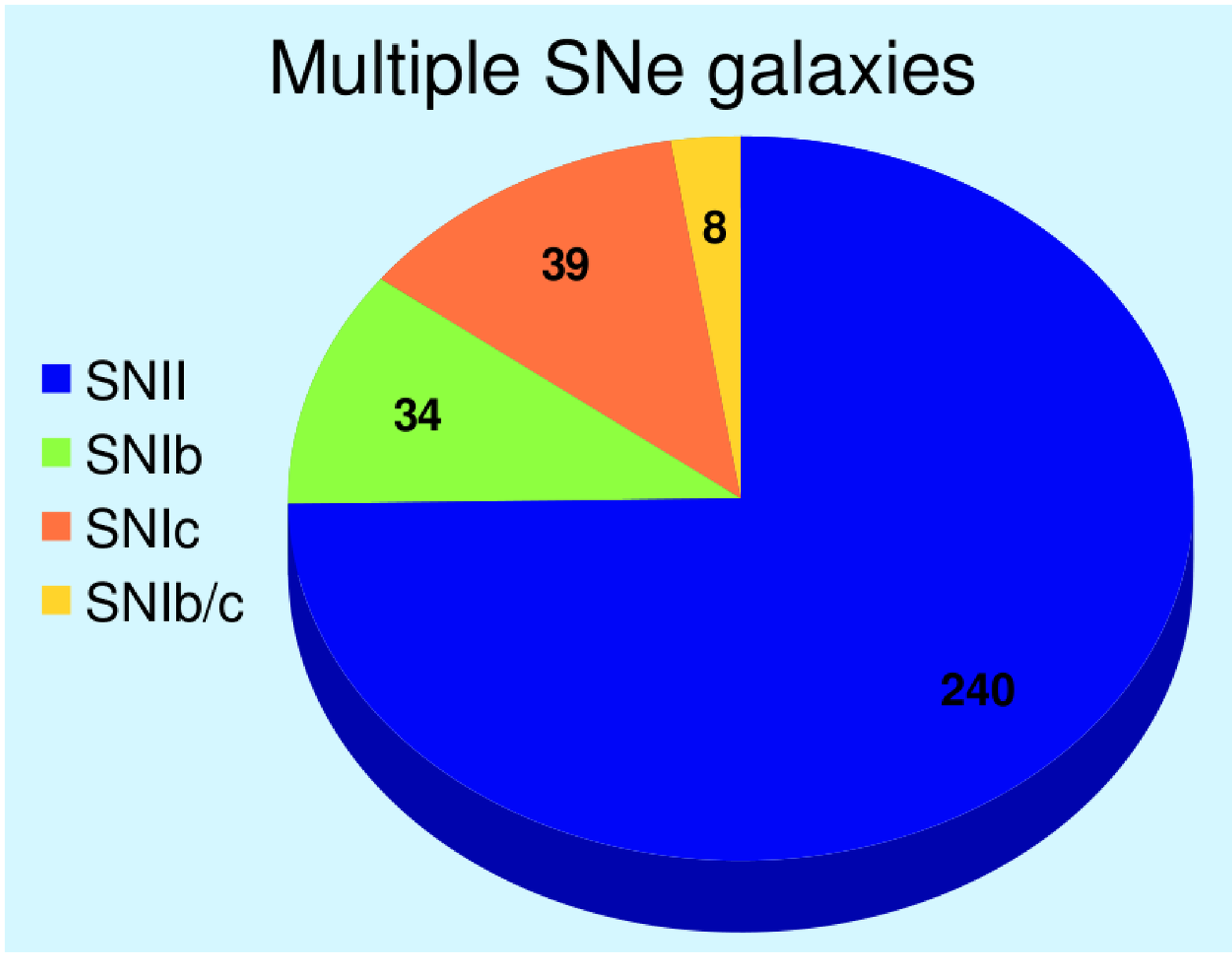}
\caption{Pie charts showing the different CC SN fractions of events within
single (top) and multiple (bottom) SN hosts.}
\end{figure}
\indent In fig. 5 we show the ratio of SNIbc to SNII as a function of number of SNe
within each host galaxy. The ratio increases with number of
hosted SNe. However, in the small number of galaxies with $\ge$4 SNe the
ratio drops dramatically. 
In fig. 6 the SNIbc sample is further split into sub-types: Ib and Ic,
and we show how their relative rates to each other, but also to SNII, change
with number of SNe per host. Here we find that the increase in the SNIbc to SNII ratio is
dominated by an increase of SNIb. While the SNIc to SNII ratio shows a
small increase, the increase in the number of SNIb to SNII is much more
significant. In single SN hosts this ratio (SNIb to SNII) is 0.072$\pm$0.011, while
it is 
0.140$\pm$0.030 in multiple SN hosts. The differences between these distributions is
significant at the 2.2 sigma level. In fig. 6 we do not plot the distributions
for galaxies host to $\ge$4 SNe due to low number statistics which would
complicate the plots. However, the values are: Ib/Ic = 2/0, Ib/II = 2/33,
and Ic/II = 0/33; these distributions are what we see in the last bin within
fig. 4. We will discuss this apparent lack of SNIc in the most multiple hosts
in section 5.2.
\begin{figure}
\includegraphics[width=9cm]{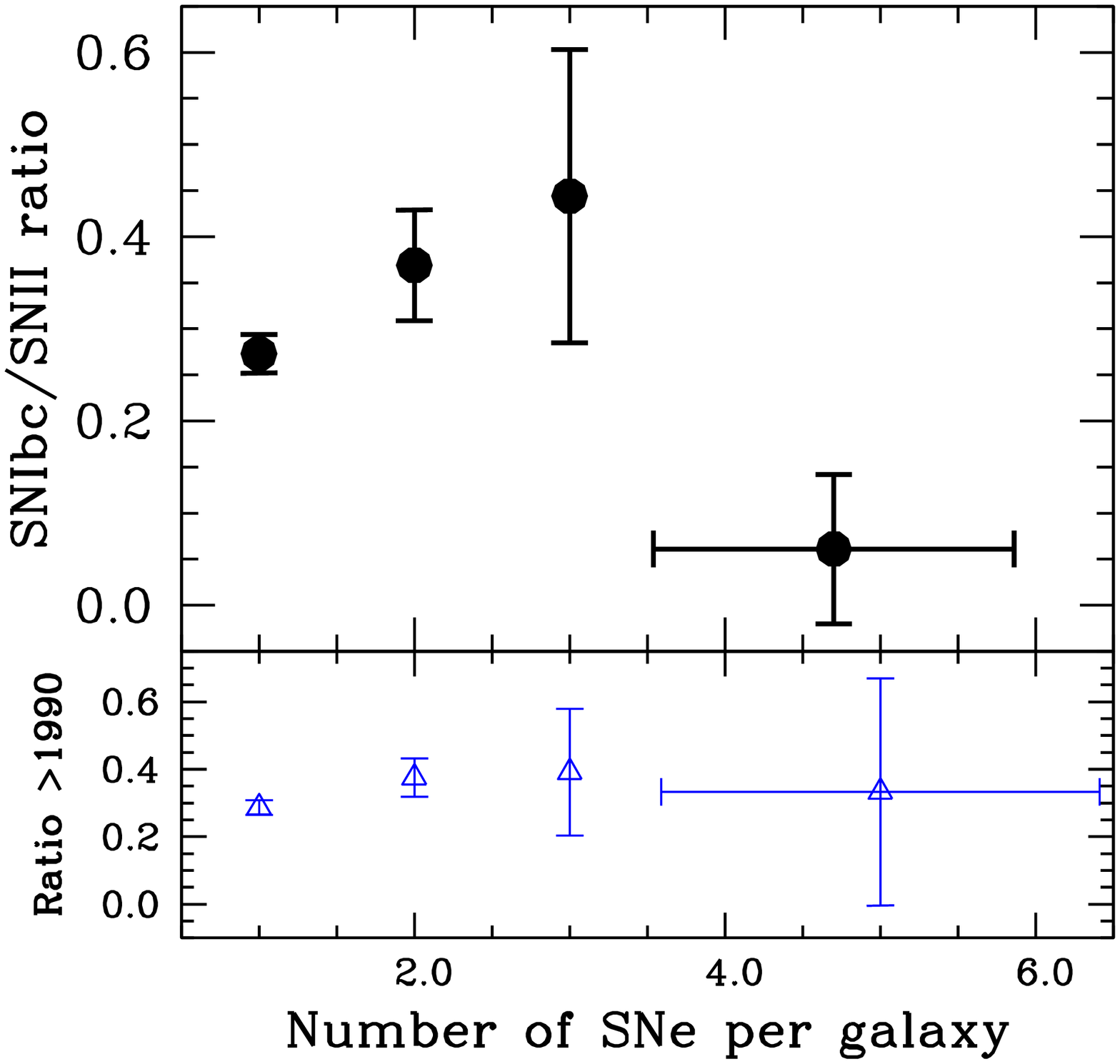}
\caption{SNIbc to SNII ratio as a function of host galaxy SN multiplicity. 
\textit{Top panel:} the full sample. The bin for galaxies with $\ge$4 SNe has a weighted
mean number of SNe of 4.7. Error bars on the ratios are
calculated using Poisson statistics (if the number of SNe within a
distribution is less than 100, then we use \citealt{geh86} to estimate
errors). The x-axis errors bars on the last point for $\ge$4
SNe show the standard deviation on the mean number of SNe within each
galaxy within the bin. \textit{Bottom panel:} a sub-sample containing only SNe
discovered after the year 1990 (see section 4.2 for more discussion). Here the bin for galaxies with $\ge$4 SNe has a weighted
mean number of SNe of 5.}
\end{figure}
\begin{figure}
\includegraphics[width=9cm]{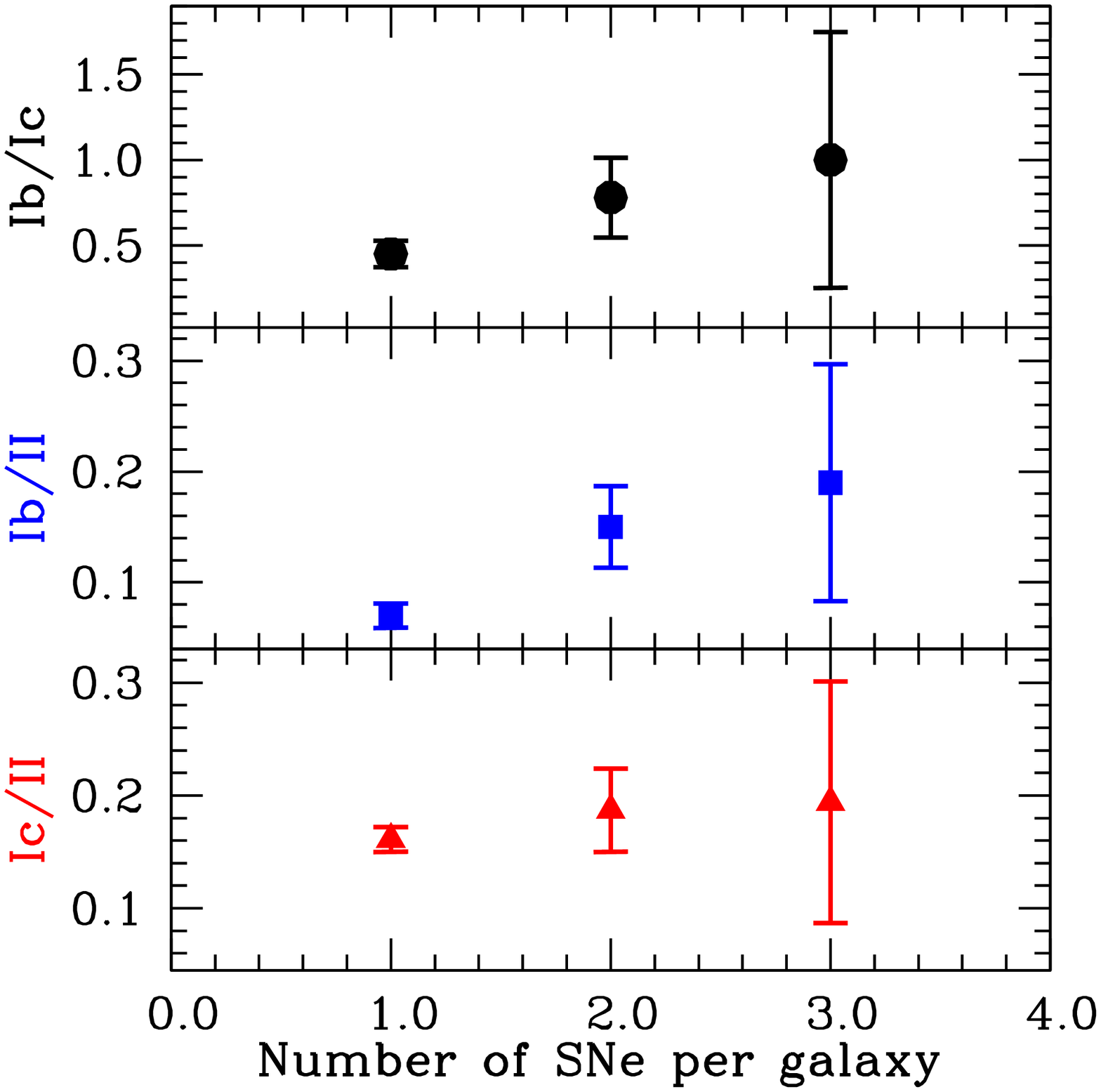}
\caption{CC SN ratios as a function of SN multiplicity of host
galaxies. In the top panel the SNIb/SNIc ratio is plotted, in the middle the
SNIb/SNII ratio, and in the bottom panel the SNIc/SNII ratio. Error bars are
calculated using Poisson statistics (if the number of SNe within a
distribution is less than 100, then we use \citealt{geh86} to estimate
errors).}
\end{figure}

\subsection{Preferential multiplicity: SNe stick together!}
In the previous section we discussed the possibility that host galaxy ratios of SN types change with
the number of SNe reported within galaxies. We now turn our attention to
investigating whether those SNe within multiple SN hosts are simply randomly
distributed throughout the overall host galaxy sample, or whether the
initial occurrence of a SN of one type increases the probability of
occurrence of another SN of the same type.
To test this hypothesis we proceed as follows. First, we concern ourselves
with separating 
SNe into CC and SNIa. For each galaxy that has been host to 2 SNe we calculate a SNIa to CC
ratio. This builds up \textit{observed} numbers of galaxies, as a fraction of
the total number of galaxies within that sample, that have been host to 2 SNIa
and 0 CC, 0 SNIa and 2 CC, and 1 of each. We then calculate the
\textit{expected} galaxy fractions for each distribution using the SNIa to CC ratio for all
galaxies host to 2 SNe (our sample of galaxies host to 2 SNe has 37\%\ SNIa,
63\%\ CC). These \textit{expected} fractions for each distribution (2 SNIa
and 0 CC, 0 SNIa and 2 CC, and 1 of each) are calculated by taking the overall
observed fraction of SNIa to CC SNe within galaxies host to 2 SNe, and
calculating the chance expectations of the fractions of galaxies that will be
host to the different possible combinations of SNIa and CC SNe.
We can then compare these \textit{expected} ratios to those
\textit{observed}. The results are shown in table 3. Next we do this same analysis
for galaxies host to 3 SNe (with possible combinations: 0 SNIa and 3 CC, 1
SNIa and 2 CC, 2 SNIa and 1 CC, and 3 SNIa and 0 CC), where the \textit{expected} ratios are calculated from the
overall SN ratios in our sample in galaxies host to 3 SNe of: 21\%\ SNIa, and
79\%\ CC. We then repeat this process but exluding SNe within
  negative T-type host galaxies. For this sub-sample the ratios of events in
  galaxies host to 2 SNe are: 33\%\ SNIa and 67\%\ CC, and in galaxies host to
3 SNe: 15\%\ SNIa and 85\%\ CC.\\
\indent Next, we concern ourselves with multiplicity of different CC types, hence
separating SNe into SNII and SNIbc.
Again we calculate the \textit{observed} and \textit{expected} ratios for 
galaxies with 2 CC SNe (with an overall fraction of 27\%\ SNIbc and 73\%\ SNII), and
the \textit{observed} and \textit{expected} ratios for those galaxies 
which have been host to 3 CC events (fractions of 31\%\ SNIbc and
69\%\ SNII).
All expected and observed ratios are shown in table 3. 
\begin{table*}
\centering
\caption{The \textit{expected} and \textit{observed} ratios of SNe within galaxies host to 2, and
host to 3 SNe. First the ratio of SNIa to CC is shown, both for the
  whole sample, then that excluding SNe within negative T-type host galaxies
  (SNe in star-forming galaxies). In the first column the
number of SNIa is listed, and in the second the number of CC. We then list the
expected percentage of events for each distribution, followed by that
observed, with the number of galaxies in each distribution listed in
brackets. We then list the \textit{expected} and \textit{observed} ratios of SNIbc to
SNII. Errors on the \textit{expected} ratios are calculated using the
relative SN numbers within each sample, assuming Poisson statistics for
distributions with more than 100 events, and using \cite{geh86} for smaller
size distributions. Errors on the \textit{observed} ratios are calculated using the
number of galaxies contributing to each distribution, again using \cite{geh86} for smaller
size distributions.} 
\begin{tabular}[t]{cccc}
\hline
Number of SNIa &  N. CC & \textit{Expected} ratio & \textit{Observed} ratio (number of galaxies)\\
\hline	
\hline
\textbf{Galaxies host to 2 SNe} & & &\\
0 & 2 & 0.40$\pm$\tiny{0.03} & 0.44$\pm$\tiny{0.06} (83)\\
1 & 1 & 0.46$\pm$\tiny{0.03} & 0.39$\pm$\tiny{0.05} (72)\\
2 & 0 & 0.14$\pm$\tiny{0.01} & 0.17$\pm$\tiny{0.04} (32)\\
\textbf{Galaxies host to 3 SNe} & & &\\
0 & 3 & 0.49$\pm$\tiny{0.13} & 0.59$\pm$\tiny{0.25} (13)\\
1 & 2 & 0.39$\pm$\tiny{0.09} & 0.23$\pm$\tiny{0.14} (5)\\
2 & 1 & 0.10$\pm$\tiny{0.02} & 0.14$\pm$\tiny{0.11} (3)\\
3 & 0 & 0.01$\pm$\tiny{0.00} & 0.05$\pm$\tiny{0.07} (1)\\
\hline
\textbf{SNe in SF galaxies}&&&\\
\textbf{Galaxies host to 2 SNe} & & &\\
0 & 2 & 0.45$\pm$\tiny{0.04} & 0.46$\pm$\tiny{0.04} (76)\\
1 & 1 & 0.44$\pm$\tiny{0.03} & 0.41$\pm$\tiny{0.06} (68)\\
2 & 0 & 0.11$\pm$\tiny{0.01} & 0.14$\pm$\tiny{0.03} (23)\\
\textbf{Galaxies host to 3 SNe} & & &\\
0 & 3 & 0.61$\pm$\tiny{0.19} & 0.65$\pm$\tiny{0.28} (13)\\
1 & 2 & 0.33$\pm$\tiny{0.25} & 0.25$\pm$\tiny{0.15} (5)\\
2 & 1 & 0.06$\pm$\tiny{0.01} & 0.10$\pm$\tiny{0.09} (2)\\
3 & 0 & 0.00$\pm$\tiny{0.00} & 0.00$\pm$\tiny{0.00} (0)\\
\hline
\hline
Number of SNIbc & N. SNII & \textit{Expected} ratio & \textit{Observed} ratio (number of galaxies)\\
\hline
\textbf{Galaxies host to 2 SNe} & & &\\
0 & 2 & 0.53$\pm$\tiny{0.05} & 0.61$\pm$\tiny{0.11} (54)\\
1 & 1 & 0.39$\pm$\tiny{0.03} & 0.31$\pm$\tiny{0.07} (28)\\
2 & 0 & 0.07$\pm$\tiny{0.00} & 0.08$\pm$\tiny{0.04} (7)\\
\textbf{Galaxies host to 3 SNe} & & &\\
0 & 3 & 0.33$\pm$\tiny{0.09} & 0.63$\pm$\tiny{0.29} (10)\\
1 & 2 & 0.44$\pm$\tiny{0.11} & 0.19$\pm$\tiny{0.15} (3)\\
2 & 1 & 0.20$\pm$\tiny{0.04} & 0.13$\pm$\tiny{0.13} (2)\\
3 & 0 & 0.03$\pm$\tiny{0.01} & 0.06$\pm$\tiny{0.10} (1)\\
\hline
\hline
\end{tabular}
\end{table*}
We find that SNe of specific types appear to `prefer' to occur in galaxies containing
a SN of the same type. This is seen in the fact that the \textit{observed} ratios of SNe
are all larger than those \textit{expected} for galaxies which have been host to one specific
SN type (initially just between Ia and CC, then between Ibc and II within the
CC sample). In general the opposite is also true; in galaxies host to
multiple SNe of different types the observed ratio is generally lower than the expected one. 
The number of galaxies contained within each distribution is shown
in brackets in the last column of table 3. We concede that these
low numbers lead to statistically insignificant differences between the
expected and observed distributions on individual ratios (as indicated
by the errors on each ratio). However, given that \textit{every
distribution of SN types which contain only one SN type has an observed ratio
higher than that expected}, we believe that we are observing an interesting
real trend, and not simply the result of random fluctuations. Similar trends
were suggested in \cite{tho09}.\\

\subsection{Host galaxy T-types}

\begin{table}
\centering
\caption{Median and mean T-type values for SN host galaxy population separated
by the number of hosted SNe. The first two lines show the distributions for
the full sample (including galaxies host to SNIa) for galaxies host to 1 or
$\ge$2 SNe. We then list 
host T-type values for only the CC sample, for galaxies host to
1, 2, 3 or $\ge$4 SNe.}
\begin{tabular}[t]{ccc}
\hline
N SNe per galaxy &  Median T-type & Mean T-type \\
\hline	
\hline
\bf{Full sample} & &\\
1 & 3.70 & 3.16\\
$\geq$2 & 4.30 & 4.07\\
\hline
\bf{Only CC SNe} & & \\
1 & 4.00 & 4.30\\
2 & 4.40 & 4.43\\
3 & 5.10 & 4.85\\
$\geq$4 & 5.90 & 5.60\\
\hline
\hline
\end{tabular}
\end{table}

In table 4 we list the mean and median T-type \citep{dev59} values for distributions of
single and multiple SN hosts. Galaxy T-type
classifications basically follow the Hubble tuning fork: most negative values
are spherical ellipticals, while most positive values are irregulars (see \citealt{but94}).
We observe clear differences in the distributions.
In fig. 7 we show these distributions
for
single and multiple SN hosts. 
A significant difference is found that multiple SN hosts
have higher T-types than the single SN hosts: a KS-test gives
less than 0.1\%\ probability that the two T-type distributions are drawn from
the same parent population. This is due in part
to the smaller fraction of SNIa hosting galaxies within the multiple sample,
which translates into a lack of elliptical, i.e. negative T-type,
galaxies. However, if we remove negative T-types the probability of the two distributions (single and
multiple SN hosts) being drawn from the same parent population is still less
than 0.1\%. \\
\begin{figure}
\includegraphics[width=9cm]{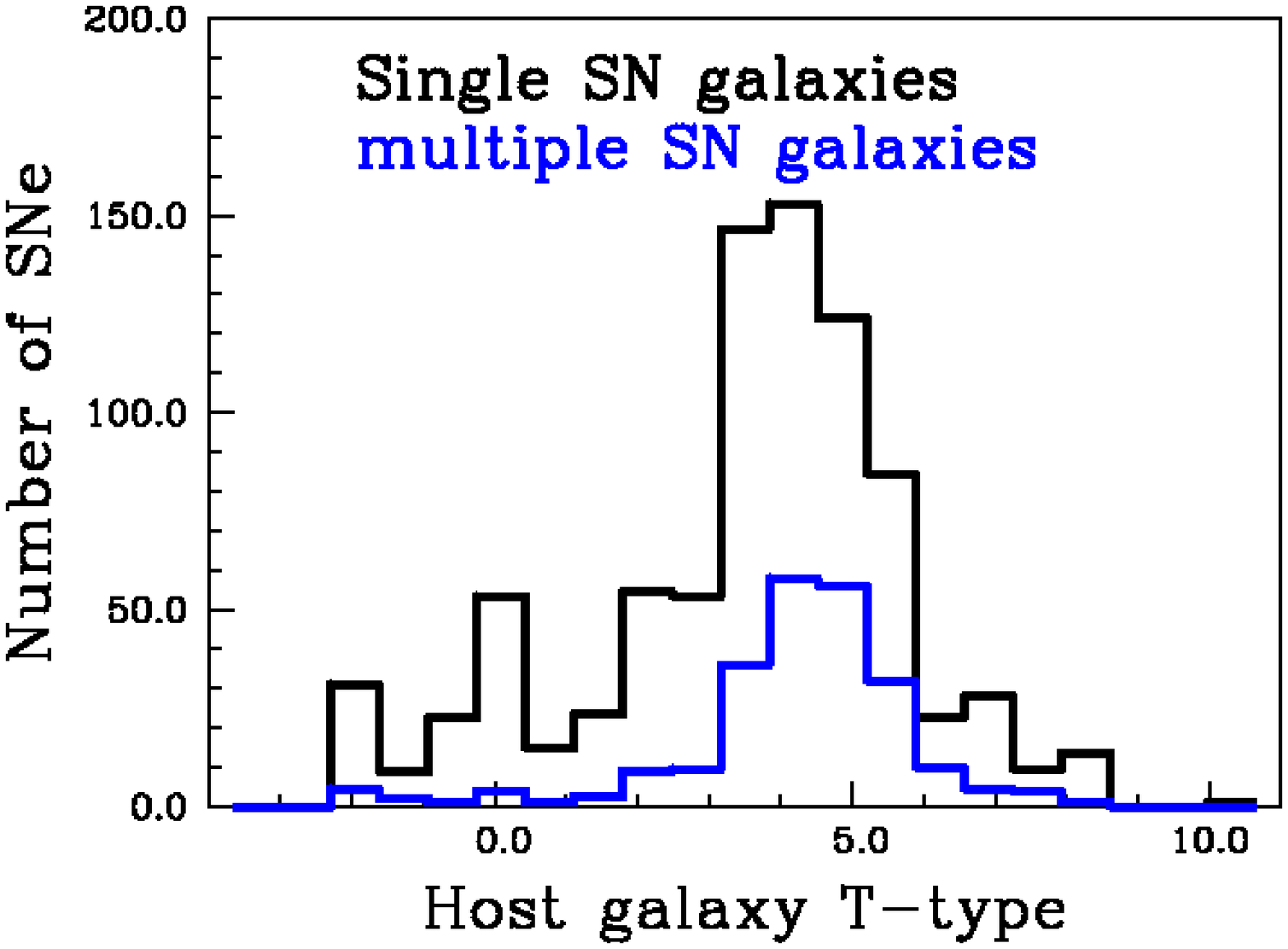}
\caption{Histograms of T-type distributions of galaxies which have hosted
1 (black), and $\ge$2 (blue) SN.}
\end{figure}
\indent In fig. 8 we remove the SNIa and show the T-type distributions of the CC sample,
separated into samples with galaxies that have hosted 1, 2, 3, and $\ge$4
SNe. There is a clear sequence of increasing galaxy T-type with SN
multiplicity. This is most clearly shown in the median T-types for each
distribution shown by the lines at at the top of the figure. We perform a
KS-test on the CC SN host T-type distributions for single and multiple SN hosts,
and find that there is less than 0.1\%\ probability of the two being drawn
from the same parent population.

\begin{figure}
\includegraphics[width=9cm]{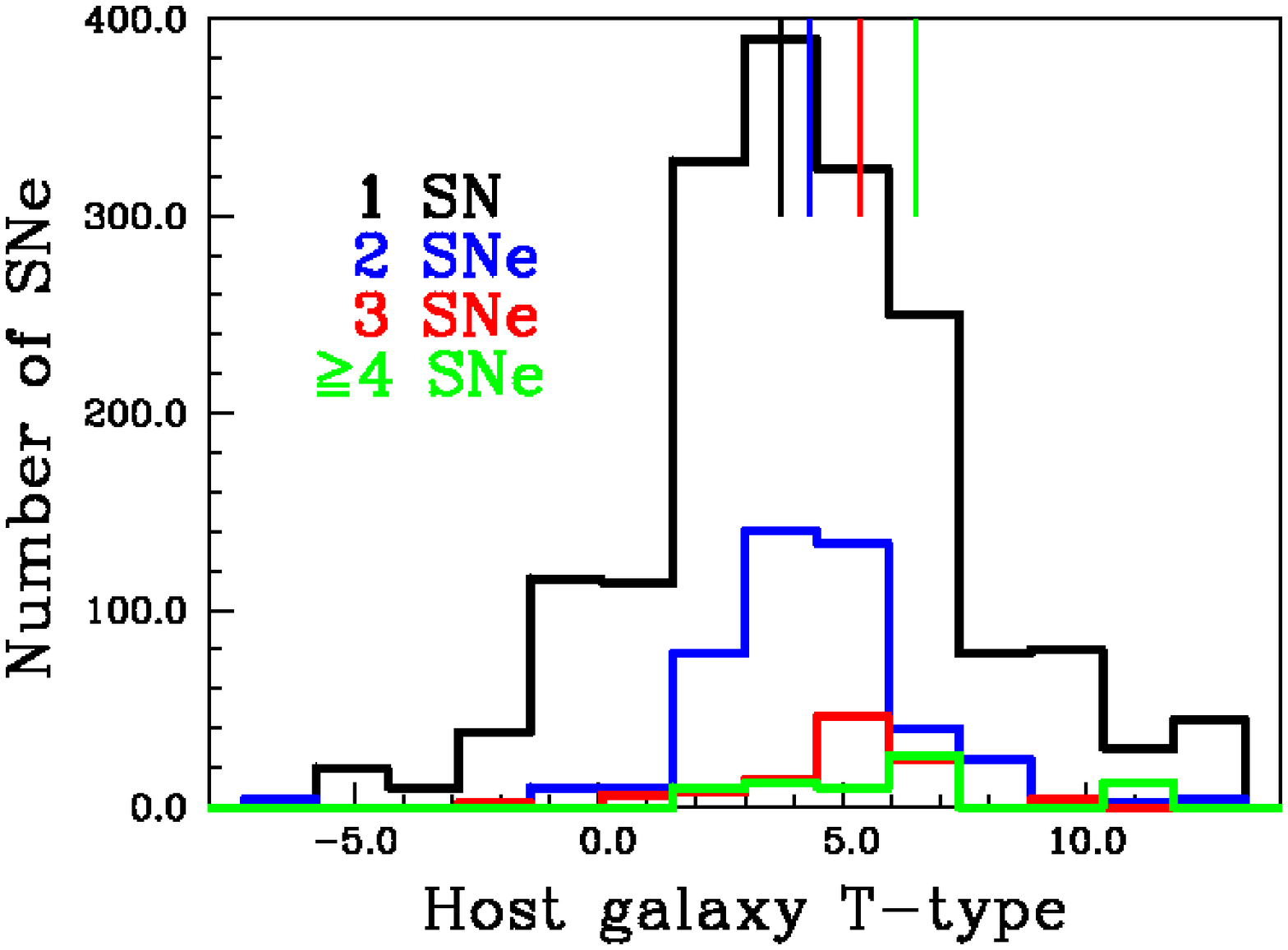}
\caption{Histograms of galaxy T-type distributions for samples which have hosted 1 (black),
2 (blue), 3 (red), and $\ge$4 (green) CC SNe. Coloured lines at the top of the
figure indicate the positions of the median T-type values for each distribution.}
\end{figure}

\section{Selection effects}

\begin{table}
\centering
\caption{Properties of galaxies, separated by number of hosted SNe. In column
1 the number of hosted SNe is listed. In column 2 the median recession
velocity of each sample is given, followed by the median galaxy $B$-band apparent
magnitude, and finally the median absolute galaxy $B$-band
magnitude. Recession velocities and apparent magnitudes are taken from the
Asiago catalogue, while absolute magnitudes are taken from the LEDA
database \citep{pat03_2}. We first list values for the full sample (including
SNIa), and then for only the CC SN sample.}
\begin{tabular}[t]{cccc}
\hline
N SNe per galaxy &  Median Vr & Median mB & Median MB \\
\hline	
\hline
\bf{Full sample} & & &\\
1 & 5485 & 14.60 & -20.74 \\
2 & 4495 & 13.46 & -21.13 \\
3 & 1221 & 11.11 & -21.04 \\
$\geq$4 & 1574 & 10.08 & -21.13 \\
\hline
\bf{Only CC SNe} & & &\\
1 & 4955 & 14.51 & -20.64 \\
2 & 4249 & 13.36 & -21.04 \\
3 & 1167 & 10.84 & -21.00 \\
$\geq$4 & 1414 & 10.08 & -21.06 \\
\hline
\hline
\end{tabular}
\end{table}

Before we discuss possible
explanations for our results, some consideration of the selection effects
within our sample is warranted. The current analysis is performed using data
compiled from a large range of sources, and hence has numerous different
selection effects convolved into one sample. The overwhelming majority of those
sources are SN searches; either professional or amateur in nature. In any search a
number of SN detection and host galaxy type selection effects will be
present. The most prominent of these can probably be summarised as
follows:\\
1) As luminosity falls with distance it is easier to detect closer SNe that
those further away.\\
2) As SNe are transient events, their detection is dependent on how frequently
one observes the same galaxy or area of sky; i.e. the cadence of any search.\\
3) The rate of SNe within any given galaxy is dependent on the galaxy's star
formation rate (SFR) (e.g. \citealt{bot11}), and in addition for the case of
SNIa, 
its mass \citep{sul06}.\\
Historical searches have generally prioritized SN detection over SN and or
host galaxy completeness. To maximise SN discovery observers have searched nearby,
low inclination face-one, large star-forming galaxies, as frequently as possible. Therefore, we would
expect these biases to arise as selection effects within our sample.\\
\indent In table 5 we list 3 host galaxy properties for each of our single and
multiple SN host samples: galaxy recession velocity,
galaxy apparent $B$-band magnitude, and galaxy absolute $B$-band
magnitude. We show these parameters for the samples including SNIa and those
without. Immediately we see that multiple SN hosts are closer to
us and therefore brighter on the night sky, in both the samples
including and excluding SNIa. These differences are
statistically significant, with KS-test probabilities in each case being less
than 0.1\%\ that the two distributions are drawn from the same parent
population. 
As we outline above, these differences are fully expected. What is
of more interest, in relation to the physical understanding of our results, is
whether differences in host galaxy properties persist once one removes the
effect of distance. To probe this we list the median absolute host galaxy $B$-band
magnitudes of each sample. Multiple SN hosts are, 
on average, around 0.5 magnitudes brighter than their single
SN hosting counterparts. A KS-test shows that the probability of observing
this difference by chance is less than 0.1\%\ (in both cases including and
excluding SNIa). Later we discuss this luminosity difference in more detail
with respect to the SN multiplicity of host galaxies.\\

\subsection{Sub-sample 1: host galaxies with recession velocities $<$5000\kms}

A different parameter which could affect our work is the ease of detecting
different SNe types. SNIa are generally of higher luminosity than CC events
(see e.g. \citealt{li11}), meaning that they should be easier to
detect. It has been shown that
while there exists a low-luminosity tail in the SNII population, their mean
peak luminosities are not significantly different from those of SNIbc
\citep{li11}. Hence, we do not expect any significant selection effect
favouring the detection of SNIbc over SNII. To test whether differences in 
SN luminosities could affect our results we repeat our
analysis but only including SNe which have host galaxies with recession
velocities less than 5000\kms. This `nearby' sample has 1123 SNe within 938 galaxies.
If there were any significant bias within our
results dependent on SN luminosity, we would expect these to be removed once
selecting only those SNe within nearby galaxies. With respect to the
relative SN fractions within single and multiple SN hosts we still observe the trends
presented above: 1) the SNIa to CC ratio (within SF
galaxies) \textit{decreases} in multiple SN hosts (0.427$\pm$0.036 in
single, and 0.336$\pm$0.045 in
multiple SN hosts), 
and 2) the SNIbc to SNII ratio \textit{increases} in multiple SN hosts (0.295$\pm$0.031 in
single, and 0.359$\pm$0.054 in multiple SN hosts).\\

\subsection{Sub-sample 2: SNe discovered after 1990}
Initially SNe were solely classified spectroscopically as types II or I based
on the presence or absence of hydrogen in their spectra \citep{min41} and it
is was not until the 1980s \citep{whe86} that the three separate type I classes; Ia, Ib
and Ic were definitively catalogued. Hence, given that the longer a galaxy has
been monitored, the higher the likelihood of multiple SN discoveries therein,
one may speculate that one may bias their results due to
incomplete SN classifications at earlier historical times. To test this we
re-analyse our sample only including SNe that were discovered
after 1990.
We find that the results are almost completely consistent with those above:
the SNIa to CC SN ratio still decreases with host galaxy SN multiplicity
(0.859$\pm$0.041 in single, and 0.558$\pm$0.059 in multiple SN hosts), and
the SNIbc to SNII ratio still increases (0.287$\pm$0.022 in single,
and 0.375$\pm$0.056 in multiple SN hosts).\\ 
\indent One of the anomalies of the above results is that
while the SNIbc to SNII ratio increases as a function of SN multiplicity, in
the last bin of SNe in galaxies of $\ge$4 detected SNe (top panel of fig. 5), 
the ratio drops dramatically. As shown in the bottom panel of fig. 5, 
in the sub-sample we analyse here the ratio within
this final bin is now more consistent with that for other multiple (2 or 3
SNe) SN hosts.\\ 

\section{Discussion}
There are two main parameters which will affect the nature of SNe
in any given stellar population (i.e. galaxy). The first is that of
different progenitor properties; age, metallicity and binarity. The second is the nature
of the star-formation; binary fraction, the IMF, and its continuous or
episodic form. The combination of these parameters will then give rise to
the relative SN fractions we observe in galaxies host to different
numbers of SNe.\\
\indent We have seen that SNIa are relatively less abundant than CC SNe within
multiple SN hosts, and also that when 1 SNIa (or 1 CC) is found within a galaxy the
probability of the next being of the same class is higher than that
expected by chance. These observations can be explained by a progenitor
age difference between that of relatively low mass progenitor SNIa, and the
high mass progenitors of CC SNe. If we assume that this explanation is valid
then this constrains the nature of SF within, in particular, the multiple SN
hosts. 
If SF within galaxies were of a continuous nature, and had been so
for a period of time longer than the time for the majority of SNIa from the initial epoch
of SF to explode (i.e. the longest characteristic SNIa delay time: time between epoch of SF and that
of explosion;
e.g. $>$2.4 Gyr, \citealt{mao11}), then we
would expect that the SNIa to CC ratio to be similar in single or multiple
SN hosts. However, given that the fraction of SNIa goes down in multiple SN
hosts, this tells us that the dominant SF within those hosts is of a shorter
timescale. While this does not provide strong constraints on the duration of
SF episodes within galaxies, significant claims have also been published for
`prompt' SNIa delays times (see e.g. \citealt{man06}), with SNIa being
produced on timescales as short as $<$420 Myr \citep{mao11}. Hence, if a
significant fraction of SNIa were produced by progenitor systems with these
short delay times, this would limit the duration of SF episodes within these
galaxies to be less than this characteristic age.\\  
\indent There is a suggestion that the fraction of SNIbc increases with respect to SNII as a function of SN
multiplicity. 
We also find that within multiple CC SN hosts there is a preference for SNII or SNIbc to be found together. This
therefore hints at progenitor characteristics playing a role in CC SN
multiplicity.\\ 
\indent The question of which progenitor characteristics drive the
differences we see between SNII and SNIbc is a currently debated
topic. There is strong evidence that the dominant SNIIP events arise from
red-supergiant progenitors in the mass range
8-16\msun\ (\citealt{sma09}, although see \citealt{wal12} for
  extending this mass range upwards). However, it is still unclear which is the
dominant progenitor characteristic that produces a SNIbc event in
place of a hydrogen rich explosion. One path is a massive single WR star
(e.g. \citealt{gas86}), which has been stripped of its envelope through
strong stellar winds. Another route is through less massive
progenitors which have their envelopes removed through binary interaction
\citep{pod92}. Observationally while no direct detection of a SNIbc
progenitor has been documented \citep{sma09b}, these events are found closer to bright
HII regions within galaxies than SNII \citep{and08,and12}, suggesting that
they arise from shorter lived, and hence more massive progenitors (also see \citealt{kel08,kel12}).\\ 
\indent Theoretically it is easier to produce a SNIbc at higher
metallicity through both single (e.g. \citealt{heg03}) and binary
\citep{eld08} models. Studies of
the global properties of host galaxies have observed that SNIbc arise in
higher metallicity hosts than than SNII \citep{pri08b,boi09,arc10}. However, investigation of
metallicities measured at the environments \textit{within}
galaxies have not found significant differences (see
\citealt{and10,sto12,san12}). Differences between the environment
metallicities of SNIb and SNIc are also presently unclear, with various claims
and counter claims in the literature \citep{and10,mod11,lel11,san12,mod12}.\\
\indent To understand the driving factors behind multiple SN hosts producing
relatively more SNIbc than single galaxies one has to combine all the
above knowledge (or lack thereof!) on progenitors, with knowledge of how host SF
properties may change with galaxy type. 
Progenitor age as the driving effect for differences
in CC SN relative ratios is much
more problematic than in the case of SNIa. Even the relatively low mass
progenitors of SNIIP will have delay times of less than a few 10s of
Myrs. Starbursts within galaxies are generally observed to have timescales
longer than this \citep{san96,cap09}. Hence, while there is the
possibility of detecting any one galaxy which is host to only e.g. SNIc due to
the very young nature of the SF, one would expect to see just as many cases
where the dominant age was that producing SNIb or SNII, implying that the
ratio of SNIbc to SNII should not change as a function of multiplicity. Therefore it is unclear
how easily we can explain our results using this argument (see \citealt{hab12}
for a more detailed discussion of this progenitor-starburst-age issue).\\ 
\indent Progenitor age can more naturally explain why, when multiplicity does
occur SNII are more often than expected found together with SNII, and that
SNIbc are more often than expected found with SNIbc. One can speculate that
within each one of these galaxies we are seeing SF of a specific age.
This is then
evidence for SF of a bursty, episodic nature, where at any one point the
events being produced by a certain galaxy are from a SF episode of a certain
age. 
This is also interesting as it implies that bursts of SF are galaxy
wide, and poses a provocative question on how environments with large
separations within galaxies `communicate' to produce SF of very similar
ages.\footnote{We reiterate here; while bursty, episodic SF can explain why
SNe types like to occur together, it \textit{does not}
explain the overall increase in the SNIbc to SNII ratio in multiple SN hosts.}\\ 
\indent One may speculate that in place of age, progenitor metallicity is the driving
force behind this increase in multiplicity. In section 4 we found that
multiple SN hosts are more luminous than their single hosting
counterparts. More luminous galaxies are generally found to be more metal rich
\citep{tre04}. Hence, multiple SN hosts are possibly more metal rich. This
could then be compatible with the production of a higher fraction of
SNIbc. However, the fact that this higher ratio is dominated by an increase in
the number SNIb complicates the matter as SNIb are predicted (see
e.g. \citealt{heg03,eld08}) to
arise from lower metallicity progenitors than SNIc, with observations
supporting this hypothesis (\citealt{mod12}, although see
\citealt{san12}). Importantly here there are no claims that SNIb arise from
\textit{higher} values than SNIc (or indeed observations that they have
higher abundances than SNII; \citealt{and10,sto12,san12}).\\
\indent The final parameter which could be driving our results is the nature of the
IMF within different galaxies. While changes in the IMF with stellar
population are often dismissed by the community, numerous claims exist of a
varying IMF (see \citealt{bas10} for a review). How these changes could affect
the relative fractions of SNe have been discussed previously in \cite{hab10} and
\cite{hab12}. If the IMF within a certain population is biased to produce
stars of higher mass then this may be reflected in an increasing rate of SNe
produced by higher mass stars; SNIbc.\\ 
\indent We have also found that the intrinsic nature of multiple SN hosts are
different from single SN hosts. As mentioned above multiple SN hosts tend to
be intrinsically brighter. This may then relate to an increased overall SFR
when compared to single SN hosts; more SF producing more SNe. However,
\textit{an increase in the SFR by itself is insufficient to explain a change in
the relative SN fractions}. With respect to differences in host properties, a
sequence of increasing host galaxy T-type is observed, as a function of host
multiplicity, seen in fig 8. The T-type classification runs from -6 to
+11 \citep{but94}, and overall forms a sequence of decreasing galaxy mass (hence also
metallicity; e.g. \citealt{tre04}) and increasing
specific SFR, while the absolute SFR peaks at T-types of
around 4-5 \citep{jam08}. Indeed the median T-type values for the CC host
galaxies presented in table 4, are consistent with this peak in the absolute
SFR of galaxies.

\subsection{Extreme SNe producers}
4 galaxies have been host to 6 or more discovered SNe. Here we briefly discuss these
galaxies and their SN populations in more detail. We again note; our aim is
not to attempt to explain why these galaxies have produced the largest numbers
of SNe. Our interest is to explore whether the SN populations within these
galaxies are similar to `normal' galaxies, and what we can infer from asking
these questions. 

\subsubsection{NGC 6946}
NGC 6946 holds the record for highest number of discovered SNe with 9
events. In table 6 information on each SN is given, and the
positions of each SN within the galaxy are show in fig. 9. NGC 6946 is host
to a well observed and modeled nuclear starburst \citep{eng96}, and has a
relatively high star formation rate (SFR) of around 3\msun yr$^{-1}$
(e.g. \citealt{tho09}). The galaxy is a face-on spiral with a Hubble
classification of SABcd, and is located at a distance of 6.0 Mpc,
with a heliocentric recession velocity of 40\kms\ (all values taken from
NED).\\
\indent It is
interesting that all SNe within this galaxy with definitive
classifications are SNII. Even that which has a
`I' classification has been claimed to be of SNII \citep{tam82}.
To test if this apparent bias towards SNII production can simply be explained by
statistical fluctuations we randomly draw SN types from the local rates given
by \cite{li11} 10000 times. First we do this including SNIa. 
We find a chance probability of 2.1\%\ that one would find 7 out of
7 SNII (if we assume that SN 1939C is of type II the probability of 8 from 8
is 1.3\%). This is suggestive that the SF producing the SNe within
this galaxy is related to a SF episode with an age of less than a few 10s
Myrs and not from continuous SF. This is because if the current
SF were part of a continuous cycle then we would expect to see SNe
with different characteristic progenitor delay times. When we draw only from the CC
relative fractions, this probability increases to 12.9\%\ (9.8\%\ for 8
SNII). Hence, one may speculate that all of the SNe
within NGC 6946 originate from a single episode of star formation with an age
of 10-20 Myrs ago; i.e. on longer timescales than SNIbc
progenitors 
(for stellar model predictions see
e.g. \citealt{heg03}).

\begin{table*}
\centering
\begin{threeparttable}
\caption{SNe discovered in the galaxy NGC 6946. In the first column we list
  the SN name, followed by the discovery reference in column 2. We then list the SN
  type classification, followed by the relevant reference.}
\begin{tabular}[t]{cccc}
\hline
SN name &  disc. Ref. & SN type & class. Ref. \\
\hline	
\hline
1917A & Ritchey\tnote{1} & II & \cite{bar79}\\
1939C & \cite{zwi42} & I & \cite{bar79}\\
1948B & Mayall\tnote{2} & IIP &  \cite{bar79}\\
1968D & Wild \&\ Dunlap\tnote{3} & II & \cite{bar79}\\
1969P & \cite{ros71} & ? & NA \\
1980K & \cite{wil80} & IIL & \cite{kir80,but82}\\
2002hh & \cite{li02}& IIP & \cite{fil02} \\
2004et & \cite{zwi04}& IIP & \cite{zwi04,fil04}\\
2008S & \cite{arb08} & IIn & \cite{sta08} \\
\hline
\hline
\end{tabular}
\begin{tablenotes}
\item$^1$No official reference was found for this discovery, hence we simply
list the discovery author in the catalogues
\item$^2$As above
\item$^3$As above
\end{tablenotes}
\end{threeparttable}
\end{table*}

\begin{figure}
\includegraphics[width=9cm]{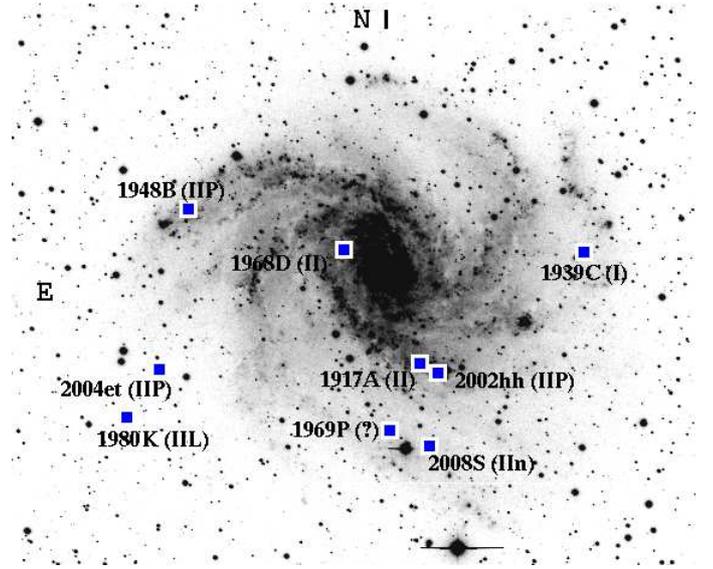}
\caption{The positions of the 9 SNe within the galaxy NGC 6946. This image is
a negative $R$-band image obtained with the Isaac Newton Telescope
(INT). These data were initially analysed in \cite{and08}. The
orientation is indicated on the image, and the scale bar (to the right of `N')
shows 20 arc seconds.}
\end{figure}

\subsubsection{Arp 299}
Second place in the list for SN producers goes to the interacting galaxy pair
(NGC 3690 and IC 694): Arp 299.  In table 7 information on each SN is
given. The positions of the SNe within this system are shown in fig. 10.\\
\indent Arp 299 is by far the most distant of the extreme SN producers discussed here
at 43.9 Mpc, with a heliocentric recession velocity of 3121\kms\ (both taken
from NED).
Due to the merging of the two components, Arp 299 is going through an episode
of intense merger-driven burst of SF (e.g. \citealt{wee72,rie72,geh83}), and
various young starburst regions have been identified by
e.g. \cite{alo00}. Through radio observations (\citealt{nef04}, see also
\citealt{per09,ulv09,rom11,bon12}) 
the nucleus of component IC 694 has been dubbed a `SN
factory' with an estimated CC SN rate of 0.1-1 events per year. 
The CC SN rate of the whole system has been estimated to be 1.5-1.9
  SNe per year \citep{mat12}. Almost
all of these SNe are undetected at optical wavelengths (see
e.g. \citealt{mat12}) due to the huge
extinction present in the nuclear regions \citep{alo00}.\\ 
\indent \cite{and11} discussed this galaxy and its SNe in detail, speculating
that both the relative numbers of SNIb and SNIIb to other SNII events, plus
their radial positions pointed to either progenitor age or IMF effect being at
play to produce the SNe types and their distribution within the system.
Arp 299 has been host to a range of CC SN types, but no SNIa. The probability
of no SNIa by chance is 16\%. Perhaps interesting is that no SNIc are detected
in this galaxy either, although the chance probability of this is only
25\%. \cite{and11} speculated that if the CC SNe within this galaxy
are all associated with a recent, interaction driven star burst, then we may
be observing the galaxy at such a time where all the stars destined to explode
as SNIc have already done so, due to their possible shorter lifetimes. 

\begin{table*}
\centering
\begin{threeparttable}
\caption{SNe discovered in the galaxy Arp 299. In the first column we list
  the SN name, followed by the discovery reference in column 2. We then list the SN
  type classification, followed by the relevant reference.}
\begin{tabular}[t]{cccc}
\hline
SN name &  disc. Ref. & SN type & class. Ref. \\
\hline	
\hline
1992bu & \cite{van94} & ? & NA \\
1993G & \cite{tre93} & IIL & \cite{fil93_2,tsv94}\\
1998T & \cite{li98} & Ib & \cite{li98} \\
1999D & \cite{qiu99} & II & \cite{jha99} \\
2005U & \cite{mat05} & IIb& \cite{mod05,leo05} \\
2010O & \cite{new10} & Ib & \cite{mat10} \\
2010P & \cite{mat10_2}& IIb& \cite{ryd10,her12}\tnote{1} \\
\hline
\hline
\end{tabular}
\begin{tablenotes}
\item$^1$Originally this event was classified as `IIb/Ib' and hence
was not part of the sample analysed in earlier sections. \cite{her12} have
recently claimed that radio observations constrain this to be of type IIb nature
\end{tablenotes}
\end{threeparttable}
\end{table*}

\begin{figure}
\includegraphics[width=9cm]{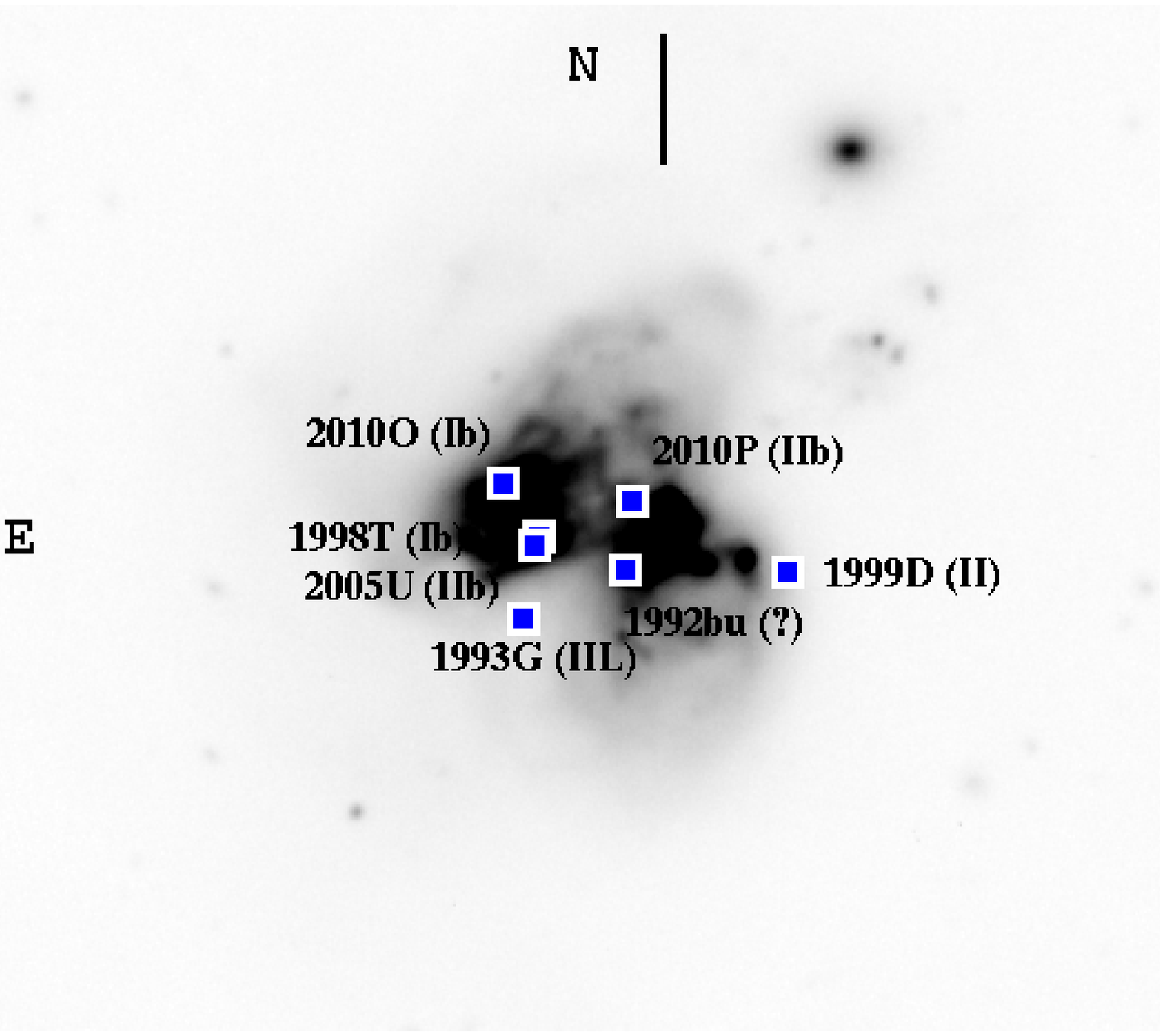}
\caption{Positions of the 7 SNe that have been discovered within Arp
  299. This image is a negative $R$-band image obtained with the Isaac Newton Telescope
(INT). These data were initially analysed in \cite{and08}. The
orientation is indicated on the image, and the scale bar (to the right of `N')
shows 20 arc seconds.}
\end{figure}

\subsubsection{NGC 4303}
As show in table 8, the 6 SNe that have been discovered within NGC 4303
(Messier 61) are all SNII. SN positions within the galaxy are shown in fig. 11. NGC 4303 is a
face-on SABbc type galaxy located at a distance of 16.5 Mpc, with a recession
velocity of 1566\kms\ (from NED). This galaxy has a high star formation rate
of $\sim$10\msun yr$^{-1}$ (\citealt{tho09}, also see e.g. \citealt{mom10}),
and has nuclear starburst activity \citep{col00}.
The detection of 6 SNe all of type II by chance has a 4.0\%\
probability if we include SNIa, and   
17.4\%\ if we draw SNe only from a CC ratio distribution. Hence, we speculate
that this is again pointing to a recent (but not too recent for SNIbc to
be detected) burst of SF dominating over long-term
continuous SF within this galaxy.\\ 

\begin{table*}
\centering
\begin{threeparttable}
\caption{SNe discovered in the galaxy NGC 4303. In the first column we list
  the SN name, followed by the discovery reference in column 2. We then list the SN
  type classification, followed by the relevant reference.}
\begin{tabular}[t]{cccc}
\hline
SN name &  disc. Ref. & SN type & class. Ref. \\
\hline	
\hline
1926A & Wolf, Reinmuth\tnote{1} & IIL & \cite{bar79}  \\
1961I & \cite{hum62} & II & \cite{pat72} \\
1964F & Rosino\tnote{2} & II\tnote{3} &  \\
1999gn & \cite{dim99} & IIP & \cite{aya99} \\
2006ov & \cite{puc06} & IIP & \cite{puc06} \\
2008in & \cite{oks08} & IIP & \cite{cha08_2} \\
\hline
\hline
\end{tabular}
\begin{tablenotes}
\item$^1$No official reference was found for this discovery, hence we simply
list the discovery author in the catalogues
\item$^2$As above
\item$^3$This originally appeared in the catalogues as type I \citep{bar84},
however given that both the IAU and Asiago catalogues now list this as type
II, we adopt this secondary classification 
\end{tablenotes}
\end{threeparttable}
\end{table*}

\begin{figure}
\includegraphics[width=9cm]{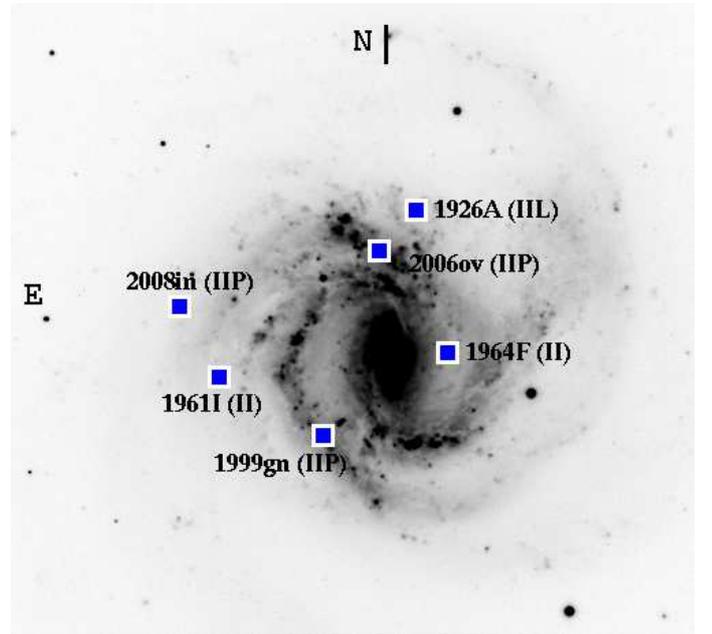}
\caption{Positions of the 6 SNe that have been discovered within NGC
  4303. This image is a negative $R$-band image obtained with the Isaac Newton Telescope
(INT). These data were initially analysed in \cite{and08}. The
orientation is indicated on the image, and the scale bar (to the right of `N')
shows 20 arc seconds.}
\end{figure}

\subsubsection{NGC 5236}
NGC 5236 (Messier 83) has been host to 6 SNe, however given that only 3 have definitive
classifications, it was not considered as part of the `$\ge$4' sample above. 
NGC 5236 is  a relatively face-on galaxy with a
morphological SABc classification, located at a distance of $\sim$7 Mpc, with
a recession velocity of 513\kms\ (NED). It is host to a nuclear
starburst (see e.g. \citealt{boh83}), and has a SFR of a few solar masses per
year (depending on the SF tracer used; \citealt{tho09}, see
\citealt{tal80} for an overview of SF in the galaxy).
The SNe are listed in table 9 and their positions within NGC
5236 are shown in fig. 12. The occurrence of long-lasting radio
emission from both SN1940B and SN1957D led \cite{ric84} to classify these as
SNII. 
Hence, if we assume these classifications we again
observe a galaxy producing an abundance of SNII explosions.

\begin{table*}
\centering
\begin{threeparttable}
\caption{SNe discovered in the galaxy NGC 5236. In the first column we list
  the SN name, followed by the discovery reference in column 2. We then list the SN
  type classification, followed by the relevant reference.}
\begin{tabular}[t]{cccc}
\hline
SN name &  disc. Ref. & SN type & class. Ref. \\
\hline	
\hline
1923A & \cite{lam36} & IIP &\cite{bar79} \\
1945B & Liller\tnote{1} & ? & NA \\
1950B & Haro\tnote{2} & ?& NA  \\
1957D & Gates\tnote{3} & ?& NA  \\
1968L & \cite{ben68}  & IIP & \cite{woo74}  \\
1983N & Evans\tnote{4} & Ib & \cite{ric83} \\
\hline
\hline
\end{tabular}
\begin{tablenotes}
\item$^1$No official reference was found for this discovery, hence we simply
list the discovery author in the catalogues
\item$^2$As above
\item$^3$As above
\item$^4$As above
\end{tablenotes}
\end{threeparttable}
\end{table*}

\begin{figure}
\includegraphics[width=9cm]{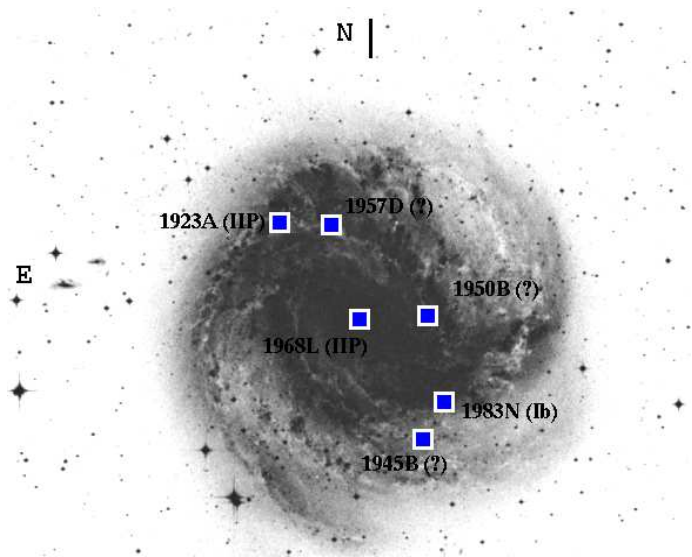}
\caption{Positions of the 6 SNe that have been discovered within NGC
  5236. This is a negative $r$-band image downloaded from the Canadian
  Astronomy Data Centre website:
  http://www4.cadc-ccda.hia-iha.nrc-cnrc.gc.ca/dss/. The
orientation is indicated on the image, and the scale bar (to the right of `N')
shows 20 arc seconds.}
\end{figure}

\subsection{The relative lack of SNIc in highly multiple host galaxies}
Not only have zero SNIc been reported in any of the 4 galaxies considered above,
but none have been documented in any galaxy which has been host to $\ge$4 
classified SNe. Drawing from the local CC rates \citep{li11}, we find that
this absence is significant, with only a 0.05\%\ chance probability (no SNIc
out of 46 SNe detected in galaxies host to $\ge$4 SNe). Even if we only
consider the 4 galaxies above, we find that there is only a 0.6\%\ chance
probability of detecting no SNIc if the galaxies were intrinsically
producing relative CC SN rates consistent those in the local
Universe. This is a puzzling result, especially given the discussion above on how
progenitor age and metallicity effects could be driving the overall higher 
relative rates of SNIbc to SNII within multiple SN hosts. One possible
explanation could be that if the progenitors of SNIc are of high mass,
e.g. $\ge$25-30\msun, then, each SF episode only produces a few explosions
on very short timescales, and hence the probability of detecting numerous SNIc
from the same SF episode is lower than that of lower mass progenitor SNII and
SNIb (although we note that in terms of the rate of their detection SNIb are
rarer events than SNIc; e.g. \citealt{li11}).\\ 
\indent While the required detailed modeling of
SF and progenitor properties to gain further answers to these results are beyond
the scope of this paper, the possibility of gaining knowledge on both is
intriguing, especially when larger statistics for this type of study are available.

\section{Conclusions}
Using a compilation of the vast majority of SNe discoveries to date, we have
investigated whether the relative fractions of SNe change with SN multiplicity
of host galaxies. We find that the
SNIa to CC ratio decreases as a function of host multiplicity, which is to be
expected given the difference in delay times between the two types of SN. 
There is a suggestion that the SNIbc to SNII ratio rises with increasing host galaxy SN
multiplicity, while in the most prolific SNe producers this trend is reversed
and we see an abundance of SNII. 
The initial increase is dominated by an increase in the number of SNIb;
a result which is difficult to interpret given the current consensus on both
progenitor characteristics and SF processes.\\
\indent We also find that within multiple SN hosts, SNe of the same types (i.e. SNIa, SNII
or SNIbc) appear to be found more often together than would be
expected by chance. 
If these trends are real then this would constrain SF to be episodic
and bursty in nature, rather than a continuous process integrated over the
lifetime of each galaxy.

\begin{acknowledgements}
The annoymous referee is thanked for constructive comments which
  improved the manuscript.
We thank Francisco Forster, Santiago Gonzalez-Gaitan, Stacey Habergham, and
Phil James for useful discussion which has greatly improved the content of
this work.
J. P. Anderson acknowledges support by CONICYT through
FONDECYT grant 3110142, and by the Millennium Center for
Supernova Science (P10-064-F), with input from `Fondo de
Innovación para la Competitividad, del Ministerio de
Economía, Fomento y Turismo de Chile'. This research
has made use of the NASA/IPAC Extragalactic Database (NED) 
which is operated by the Jet 
Propulsion Laboratory, California
Institute of Technology, under contract with the National Aeronautics and
Space Administration and of data provided by the Central Bureau for
Astronomical Telegrams. 
We acknowledge the usage of the HyperLeda database (http://leda.univ-lyon1.fr).
\end{acknowledgements}

\bibliographystyle{aa}
\bibliography{Reference}

\end{document}